\pdfoutput=1
\documentclass[conference,letterpaper]{IEEEtran}
\usepackage[letterpaper, left=1in, right=1in, bottom=1in, top=0.75in]{geometry}
\usepackage{enumitem}

\makeatletter
\newcommand{\linebreakand}{%
  \end{@IEEEauthorhalign}
  \hfill\mbox{}\par
  \mbox{}\hfill\begin{@IEEEauthorhalign}
}
\makeatother

\IEEEoverridecommandlockouts
\usepackage{cite}
\usepackage{amsmath,amssymb,amsfonts}
\usepackage{algorithmic}
\usepackage{graphicx}
\usepackage{textcomp}
\usepackage{xcolor}

\def\BibTeX{{\rm B\kern-.05em{\sc i\kern-.025em b}\kern-.08em T\kern-.1667em\lower.7ex\hbox{E}\kern-.125emX}}
\begin{document}

\title{Thermal Analysis of PEM Fuel Cell and Lithium Ion Battery Pack in Confined Space}

\author{
\IEEEauthorblockN{Ashita Victor}
\IEEEauthorblockA{\textit{ Department of Electrical and Electronics Engineering} \\
\textit{Ramaiah Institute of Technology}\\
Bangalore, India \\
ashita.victor@gmail.com}
\and
\IEEEauthorblockN{Abhay Shankar Jha}
\IEEEauthorblockA{\textit{ Department of Electrical and Electronics Engineering} \\
\textit{Ramaiah Institute of Technology}\\
Bangalore, India \\
jhaabhay10@gmail.com}
\and
\IEEEauthorblockN{Janamejaya Channegowda}
\IEEEauthorblockA{\textit{ Department of Electrical and Electronics Engineering} \\
\textit{Ramaiah Institute of Technology}\\
Bangalore, India \\
jc@msrit.edu}
\and
\IEEEauthorblockN{Sumukh Surya}
\IEEEauthorblockA{\textit{Powertrain} \\
\textit{KPIT}\\
Bangalore, India \\
sumukhsurya@gmail.com}
\and
\IEEEauthorblockN{Kali vara prasad Naraharisetti}
\IEEEauthorblockA{\textit{Staff Applications Engineer} \\
\textit{ Infineon Technologies}\\
California, USA \\
swaraj.kali@gmail.com}
}

\maketitle

\begin{abstract}
Hybrid energy storage systems (HESS) have carved a niche in the industry. HESS improve the system efficiency, reduce the overall cost and increase the lifespan of the system. The proton exchange membrane (PEM) fuel cell is hybridized with Li-ion batteries (LIB) for vehicular applications, robotic applications etc. In applications which have geometrical space constraints, the temperature of the energy storage elements is influenced by convective heat transfer. In this paper the thermal analysis of the geometry of PEM-LIB hybrid system is carried out using COMSOL Multiphysics Software package for different discharge rates (C rates) of the LIB and different voltages of the PEM cell. The additional rise in temperature of the LIB pack when placed in close proximity with PEM cell was in the range of 0.03-0.6$^0$C at 4C. The cell temperature of the LIB pack increased with increase in C rate and decrease in PEM cell voltage.    
\end{abstract}


\section{Introduction}

The beneficial features of Hybrid Energy Storage Systems (HESS) has led to its incorporation in several applications ranging from large-scale applications such as hybrid vehicles microgrids to small-scale applications like unmanned ground vehicles etc [1]. HESS incorporates two or more Energy Storage Systems (ESS) such as fuel cells, batteries, supercapacitors etc. to achieve the desired performance [2,3]. For example, high discharge efficiency is provided by Li-Ion batteries, but the rate capacity effect in these batteries decreases the discharge efficiency as the load current increases. Supercapacitors on the other hand, have extremely low internal resistance, and a battery-supercapacitor hybrid may mitigate the rate capacity effect [13]. Hybridization helps in effectively improving the dynamic response of the system as well as reduces the cost incurred since the ESS are sized according to the requirement [4,5,6].

The battery-fuel cell hybrid structure is one of the most popular type of hybridization. Fuel cells offer several disadvantages such as short lifetime, high cost, inability to charge during regenerative braking, slow response to sudden dynamic changes etc [7]. Thus, hybridisation of fuel cells with batteries helps in overcoming these drawbacks and thereby improves the efficiency of the system [4]. Fuel cell-battery hybrid power systems can achieve high energy density and specific energy, thereby increasing the safe conditions of the power system. 

A popular fuel cell used for hybridization is the PEM fuel cell. It consists of an anode terminal, cathode terminal, a polymeric membrane and a catalyst. Hydrogen and air can be uniformly distributed across the anode and cathode via flow fields [9]. Polymer Electrolyte Membrane or Proton Exchange Membrane (PEM) fuel cells offer several advantages such as high power density, low in weight, low noise, low pollution etc. [8]. However, the energy efficiency of the PEM fuel cell vehicle is low when the output power enters in low and high operating regions and it cannot save energy from regenerative braking. The PEM fuel cells are always combined with other energy storage devices such as batteries and supercapacitors to overcome such disadvantages [3]. In most Electric Vehicle (EV) applications, a PEM fuel cell is coupled with the battery in order to cater to the load demand [10]. PEM-battery hybrid can also be used for Unmanned Ground Vehicle (UGV) applications [1] as well as in industrial robots such as the welding robots.

In applications like the UGVs, the PEM fuel cell and the Li-ion battery pack are kept in close proximity to each other due to geometrical space constraints. Thus, the objective of this paper is to carry out the thermal analysis of the HESS for different conditions such as different discharge rates of the battery and for different voltages of the PEM fuel cell. The temperature of the cells of the LIB pack at each of these conditions is determined with the help of Multiphysics libraries available in COMSOL Multiphysics v5.5. Three dimensional plots of the surface temperature gradient for the different conditions help in locating the hotspots in the system.

This paper is divided into the following sections: Section II discusses the methodology of implementation of the geometry of the HESS, followed by the discussion of the results obtained in Section III. A conclusion of the results obtained and the future work of this paper is discussed in Section IV.

\section{Methodology}

Unmanned Ground Vehicles (UGV), like the commercially available Jackal by Clearpath Robotics, have dimensions of 508 x 430 x 250 mm, thus if a HESS consisting of a LIB pack with three cells (each having 21mm diameter, 70mm height) and a PEM fuel cell (10cm length, 2cm wide) were to be used to power the vehicle, the space between the different Energy Storage Systems (ESS) would be confined. In such cases it would be necessary to study the effect on temperature of the individual ESS when placed in close proximity.

The simulation model of the HESS consists of a PEM (Proton Exchange Membrane) fuel cell and a pack of Li-Ion Batteries (LIB).The steps involved in creating the simulated model are as follows:
\begin{enumerate}[label=\roman*]
\item	\textit{Defining Parameters}: The first step involved in designing the simulated model is to globally define the geometrical parameters such as the length, width, diameter of the cylinder, thickness etc. and the electrochemical parameters such as the voltage, activation energy, energy density, electrode potential etc., according to the requirements of both the PEM fuel cell and the LIB. 
\item	\textit{3D Modelling}: A three dimensional model is built using the globally defined geometry variables with the CAD (Computer Aided Drawing) tools available in the software package. 
\item	\textit{Material Selection}: Once the 3D geometry is built, the different materials governing the model are selected and are applied to the respective boundaries and domains. For eg. the metal strip connecting the different LIBs are made up of copper and the domain surrounding the HESS geomtery is air.
\item	\textit{Physics}: The working of the simulated model is governed by the physics defined at the different domains. Since the  multiphysics behaviour of the model is to be analysed, both the electrochemical library and the heat transfer in solids and fluids library are selected and applied at the different domains of the geometry.
\item	\textit{Study}: A time dependant study is computed to analyse the behaviour of the system.
\item  \textit{Results}: Post processing of the results obtained allows 3D vizualization of the parameters such as temperature distribution etc.
\end{enumerate}

Fig. 1. Represents the flowchart of the methodology of simulation of HESS using COMSOL Multiphysics v5.5.

\begin{figure}[htbp]
\centering
{\includegraphics[scale=0.45]{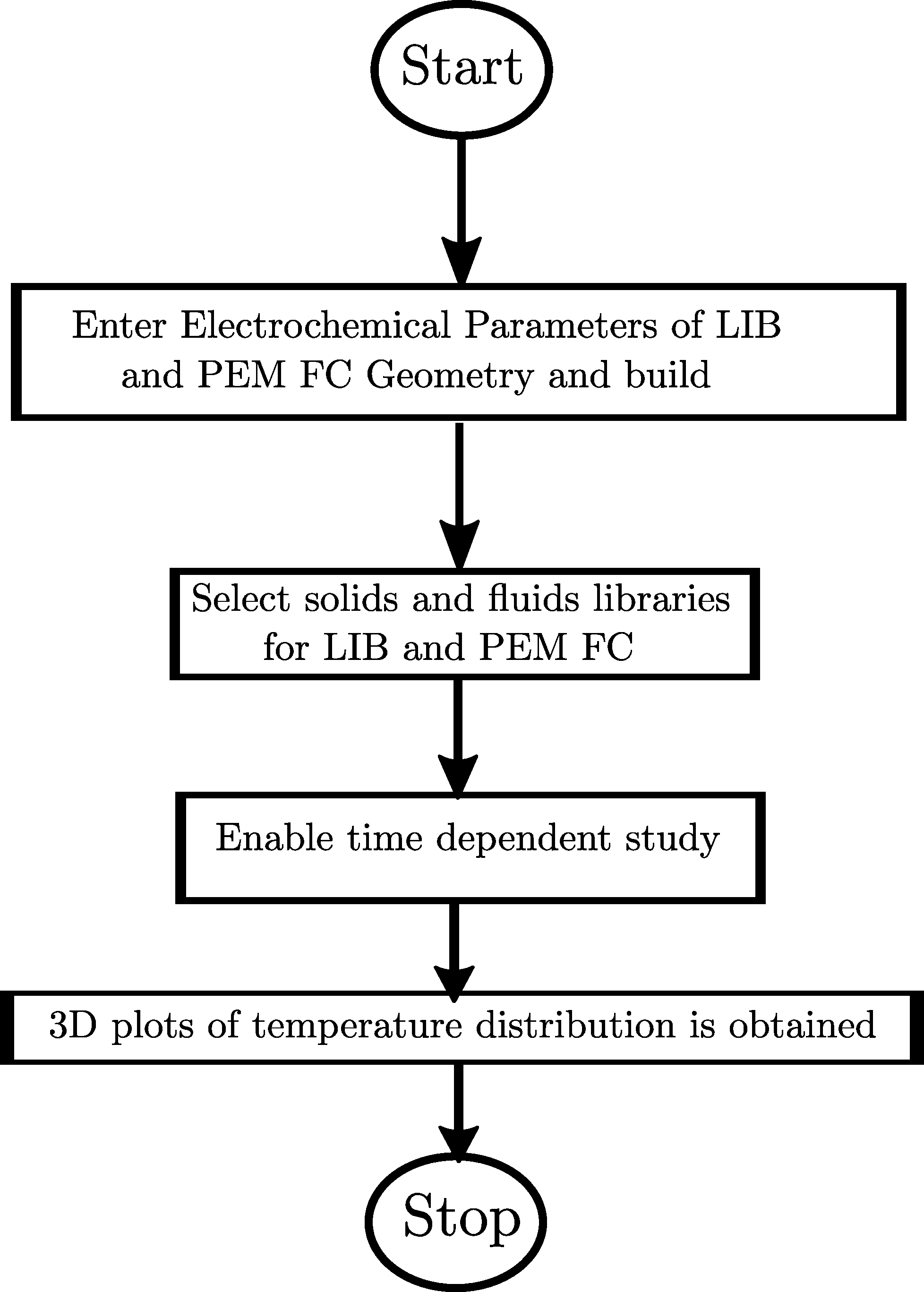}}
\caption{Flowchart of methodology of implementation}
\label{fig}
\end{figure}

In order to understand the heat transfer by convection between the closely placed PEM cell and LIB pack, the thermal analyses of the individual energy storage systems are first carried out. 

\subsection{Li-ion Battery (LIB)}

Li-ion batteries have been widely used in several applications ranging from electrical vehicles to mobile phones, as they are light in weight, have a long cycle life, high specific energy and density, high working voltage etc. A pack of six 21,700 (21mm in diameter and 70mm in height) cylindrical battery cells, where each cell has a nominal capacity of 4Ah and a nominal voltage of 3.6V, are modelled in 3D using COMSOL Multiphysics 5.5. The electrochemical behaviour of the modelled battery is governed by the Arrhenius Equation [11], 
\begin{equation}
k=Ae^{\frac{-E_{a}}{RT}}
\end{equation}

Where $k$ = rate constant\\
\hspace*{9ex}$A$ = pre-exponential factor\\
\hspace*{9ex}$E_{a}$ = activation energy\\
\hspace*{9ex}$R$ = universal gas constant\\
\hspace*{9ex}$T$ = absolute temperature in K\\

From the 3D visualization of the temperature gradient across the battery pack as seen in Fig. 2, it is clearly observed that the cells at the center of the pack (core) have a higher cell temperature ($\approx$ 4K higher) as compared to the outer cells for a discharge rate of 4C. As C rate increases from 4C to 8C, the temperature gradient across the pack increases as seen in Fig. 3.

\begin{figure}[htbp]
\centering
{\includegraphics[scale=0.36]{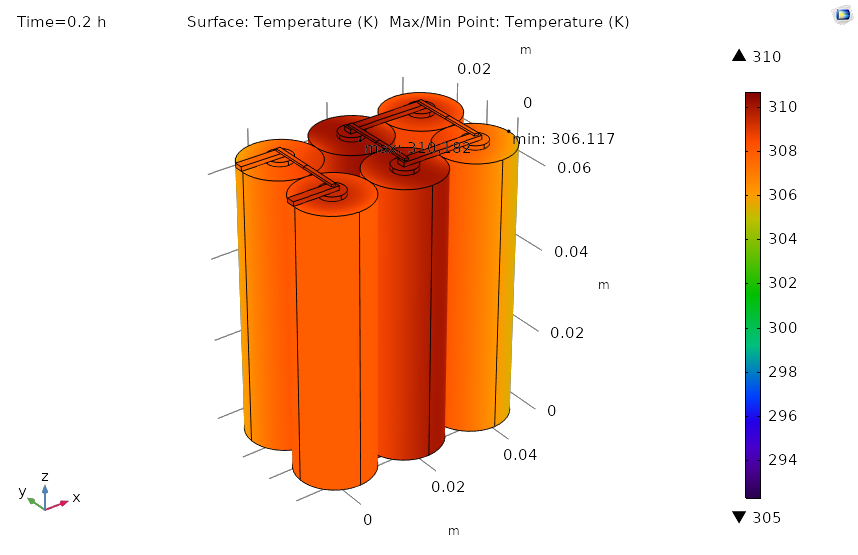}}\\
a.\\
{\includegraphics[scale=0.33]{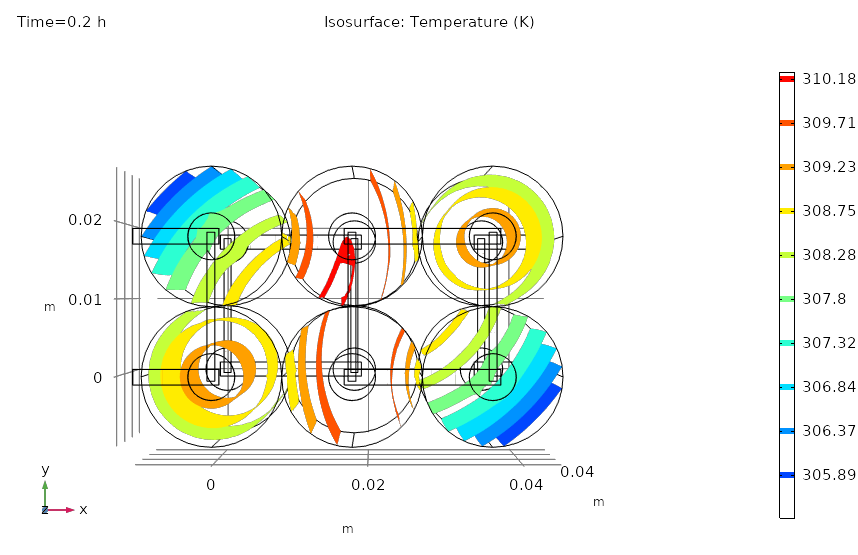}}\\
b.
\caption{Temperature distribution of Li-ion battery pack at t = 0.2h a. Surface Temperature. b. Isothermal contours }
\label{fig}
\end{figure}
\begin{figure}[htbp]
\centering
{\includegraphics[scale=0.35]{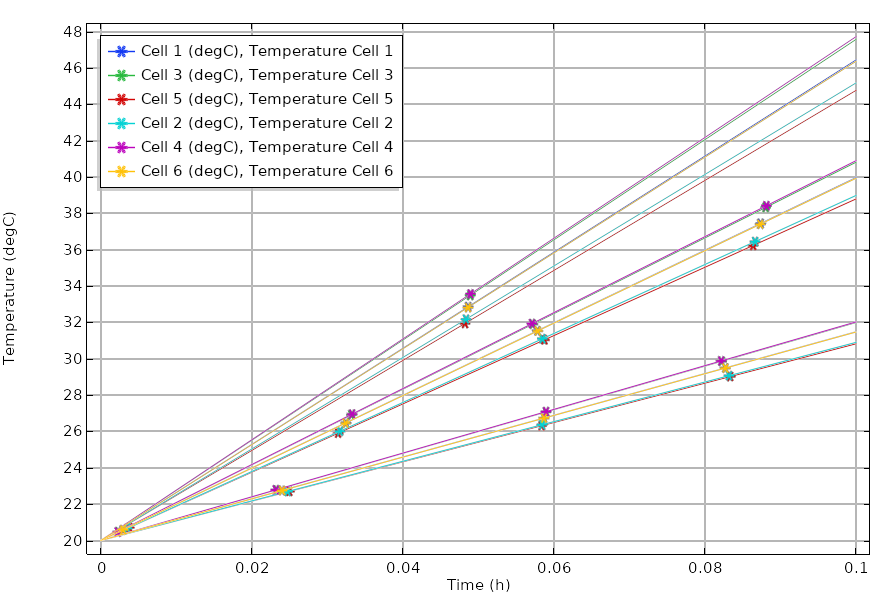}}
\caption{Cell Temperatures at 4C, 6C and 8C }
\label{fig}
\end{figure}

\subsection{Proton Exchange Membrane (PEM) Fuel Cell}

Fuel cells are devices which convert chemical energy into electrical energy by electrochemical reactions. The PEM fuel cell uses hydrogen (H$_2$) as the fuel and oxygen (O$_2$) as the oxidising agent. A polymeric membrane is used as the electrolyte. Hence, they are also referred to as Polymer Electrolyte Membrane (PEM). The 3D geometry of the PEM fuel available in COMSOL Multiphysics 5.5 is shown in Fig. 4. In order to provide good reactant transport to the cathode electrode, the cathode cover plate has holes. It is through these holes that the heat is dissipated. 

\begin{figure}[htbp]
\centering
{\includegraphics[scale=0.37]{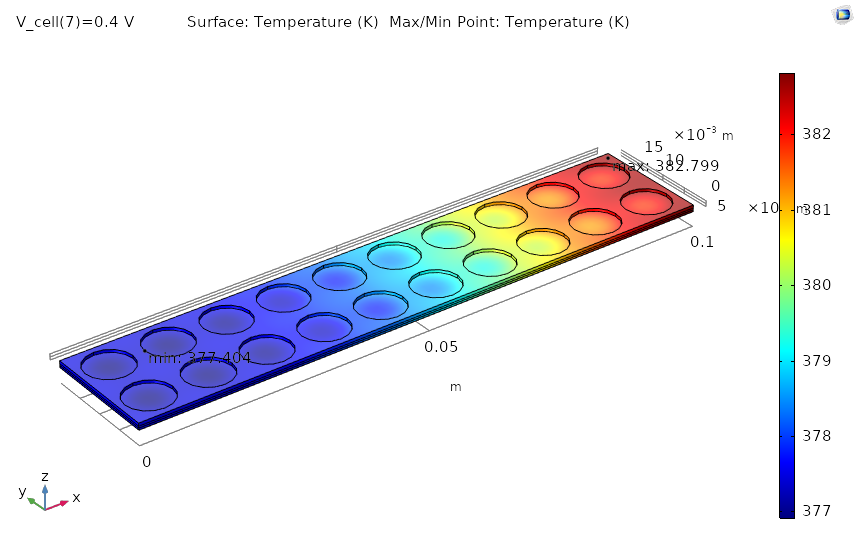}}
\caption{Surface temperature of a PEM cell}
\label{fig}
\end{figure}

The electrode kinetics of the PEM cell is governed by the Butler-Volmer equation[12],

\begin{equation}
i=i_{0}{(e^{\beta f \eta}-e^{-\alpha f \eta})}
\end{equation}
\begin{equation}
\eta=E-E_{eq}
\end{equation}

Where $i$ = faradaic current density\\
\hspace*{9ex}$i_{0}$= exchange current density\\
\hspace*{9ex}$\alpha$, $\beta$ = transfer coefficients\\
\hspace*{9ex}$\eta$ = activation overpotential\\
\hspace*{9ex}$f$ = F/RT, where F is Faraday constant, R is \hspace*{9ex}universal gas   constant, T is temperature (K)\\
\hspace*{9ex}$E$ = electrode polarization\\
\hspace*{9ex}$E_{eq}$ = equilibrium potential\\

In Fig. 4, it is observed that the temperature at the cathode electrode is higher than the temperature at the anode electrode.  When the simulation of the model is computed, it is seen that as the potential of the PEM fuel cell decreases, the average temperature of the cell increases as shown in Fig. 5.

\begin{figure}[htbp]
\centering
{\includegraphics[scale=0.34]{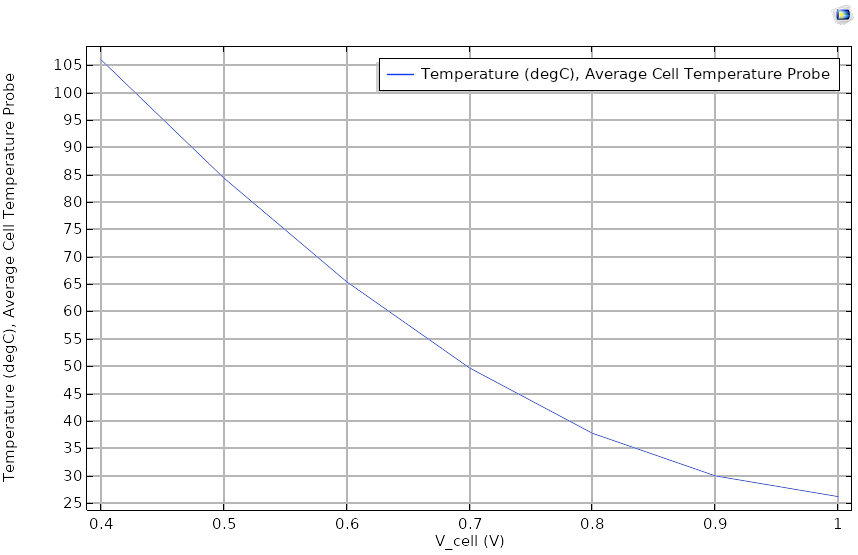}}
\caption{Average cell temperature vs. voltage plot}
\label{fig}
\end{figure}

\subsection{Hybrid Energy Storage System (HESS)}
The Hybrid Energy Storage System (HESS) consisting of the Li-Ion Battery (LIB) pack and the PEM fuel cell is considered for the study of heat propagation between the coupled storage systems as a result of confined space between them. The geometric model of the HESS in 3D is shown in Fig 6a.  A gap of 1.8 cm is considered between the LIB pack (3 cells) and the PEM cell. The entire system is surrounded by a block, which represents the air domain.

To study the heat transfer by convection, the Heat Transfer in Solids and Fluids library in COMSOL is used, where the value of the heat transfer coefficient (h) is considered as 50W/m$^2$K. The cathode cover plate of the PEM, the surfaces of the LIBs located at the ends of the pack and the air domain are considered as the heat flux domains as seen in Fig. 6b. 

\begin{figure}[htbp]
\centering
{\includegraphics[scale=0.37]{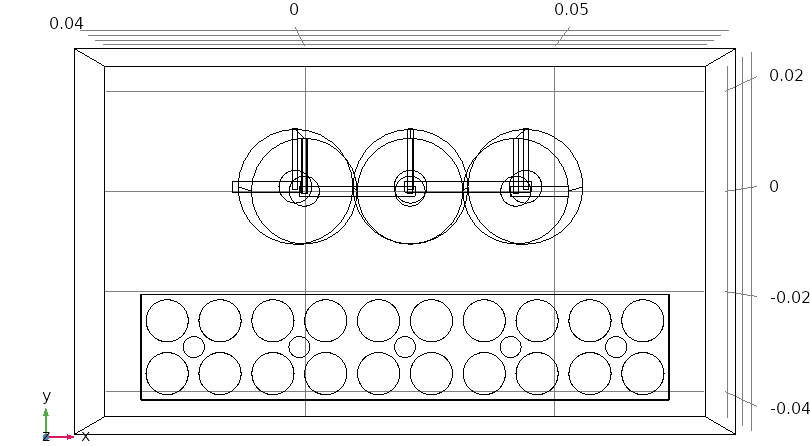}}\\
a.
\end{figure}
\begin{figure}[htbp]
\centering
{\includegraphics[scale=0.52]{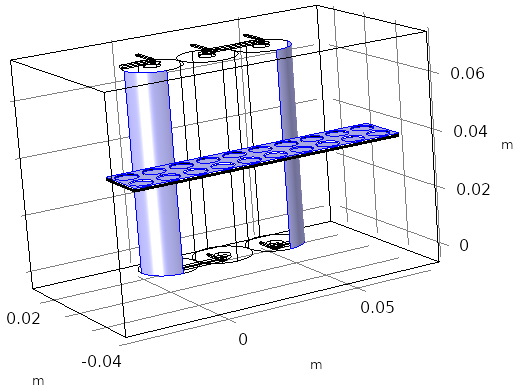}}\\
b.
\caption{Li-ion battery/PEM fuel cell hybrid a. Geometry of hybrid energy storage system b. Heat flux domains}
\label{fig}
\end{figure}

\section{RESULTS AND DISCUSSION}
A time dependent study is carried out in order to study the effect on the temperature of each cell of the LIB pack, when placed in close proximity with the PEM fuel cell at different voltages of the fuel cell and different C rates of the LIB.

In order to determine the additional rise in the temperature of the LIB cells when placed in close proximity to the PEM cell, the temperatures of the cells in the absence of the PEM cell is first determined as shown in Table 1.

\begin{table}[htbp]
\caption{CELL TEMPERATURES FOR DIFFERENT C RATES}
\begin{center}
\begin{tabular}{|c|c|c|c|c|}
\hline
\textbf{Time}&\textbf{C}&\multicolumn{3}{|c|}{\textbf{Cell Temperature($^{\circ}$C) }} \\
\cline{3-5} 
\textbf{(hr)} &\textbf{Rate} & \textbf{\textit{Cell 1}}& \textbf{\textit{Cell 2}}& \textbf{\textit{Cell 3}}\\
\hline
\text{0.1} & \text{4C} & \text{25.545} & \text{25.940} & \text{25.545}\\
\hline
\text{0.1} & \text{6C} & \text{30.119} & \text{30.762} & \text{30.119}\\
\hline
\text{0.1} & \text{8C} & \text{34.256} & \text{35.172} & \text{34.356}\\
\hline
\end{tabular}
\label{tab1}
\end{center}
\end{table}

\subsection{PEM Fuel Cell at 1V}

The temperature of the LIB cells at different C rates i.e. at 4C,6C and 8C when placed in close proximity to a PEM cell at 1V is tabulated as observed in Table 2. A 3D plot of the surface temperature, Fig. 7., at 4C shows that the maximum temperature is observed at the core of the LIB battery pack.

\begin{figure}[htbp]
\centering
{\includegraphics[scale=0.36]{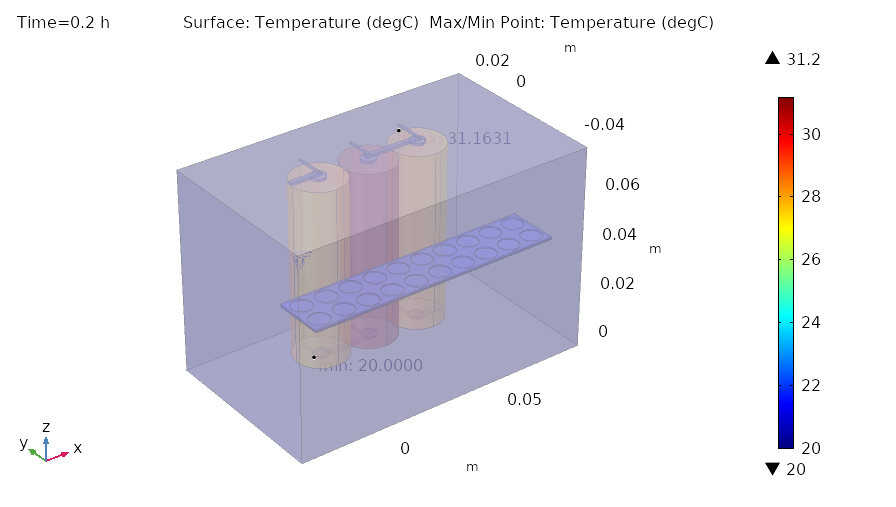}}\\
\caption{Surface Temperature at V\_cell = 1V}
\label{fig}
\end{figure}

\begin{table}[htbp]
\caption{CELL TEMPERATURES FOR DIFFERENT C RATES}
\begin{center}
\begin{tabular}{|c|c|c|c|c|}
\hline
\textbf{Time}&\textbf{C}&\multicolumn{3}{|c|}{\textbf{Cell Temperature($^{\circ}$C) }} \\
\cline{3-5} 
\textbf{(hr)} &\textbf{Rate} & \textbf{\textit{Cell 1}}& \textbf{\textit{Cell 2}}& \textbf{\textit{Cell 3}}\\
\hline
\text{0.1} & \text{4C} & \text{25.574} & \text{25.956} & \text{25.574}\\
\hline
\text{0.1} & \text{6C} & \text{30.849} & \text{31.534} & \text{30.849}\\
\hline
\text{0.1} & \text{8C} & \text{35.209} & \text{36.176} & \text{35.209}\\
\hline
\end{tabular}
\label{tab2}
\end{center}
\end{table}

On comparing the values of the temperature of the LIB cells in Table 2 and Table 1, the temperature of the cells in the presence of a PEM cell increases by $\approx$ 0.03$^{\circ}$C - 1$^{\circ}$C, for discharge rates of 4C - 8C.

\subsection{PEM Fuel Cell at 0.8V}

When the PEM fuel cell is kept at 0.8V, the temperature of each LIB cell at 4C, 6C and 8C discharge rates are determined as seen in Table 3. The temperature distribution of the HESS at a fuel cell voltage of 0.8V and at  a discharge rate of 4C of the LIB pack is visualized by the 3D plot observed in Fig.8.

\begin{figure}[htbp]
\centering
{\includegraphics[scale=0.4]{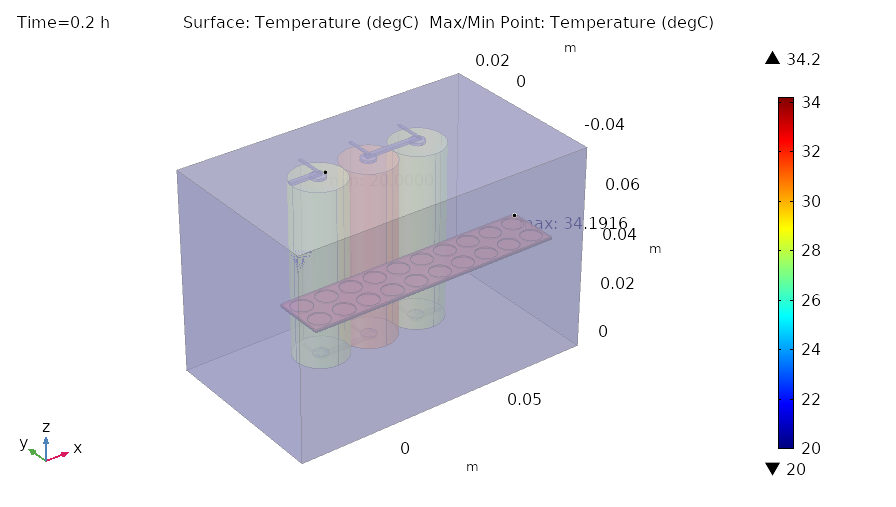}}\\
\caption{Surface Temperature at V\_cell = 0.8V}
\label{fig}
\end{figure}

\begin{table}[htbp]
\caption{CELL TEMPERATURES FOR DIFFERENT C RATES}
\begin{center}
\begin{tabular}{|c|c|c|c|c|}
\hline
\textbf{Time}&\textbf{C}&\multicolumn{3}{|c|}{\textbf{Cell Temperature($^{\circ}$C) }} \\
\cline{3-5} 
\textbf{(hr)} &\textbf{Rate} & \textbf{\textit{Cell 1}}& \textbf{\textit{Cell 2}}& \textbf{\textit{Cell 3}}\\
\hline
\text{0.1} & \text{4C} & \text{25.881} & \text{26.309} & \text{25.881}\\
\hline
\text{0.1} & \text{6C} & \text{30.711} & \text{31.471} & \text{30.711}\\
\hline
\text{0.1} & \text{8C} & \text{35.107} & \text{36.078} & \text{35.107}\\
\hline
\end{tabular}
\label{tab3}
\end{center}
\end{table}

The temperature of the cells of the LIB pack increased by   $\approx$ 0.3$^{\circ}$C – 0.9$^{\circ}$C for C rates of 4C - 8C, in comparison to the values obtained in Table 1. As the voltage of the PEM fuel cell decreases, the temperature of the PEM cell increases exponentially as observed in Fig. 5. It is this behaviour of the PEM cell that causes the temperature of the LIB pack, which is placed in close proximity to the fuel cell, to also increase in temperature as a result of convective heat transfer between the two ESS.

\subsection{PEM Fuel Cell at 0.4V}

At a constant PEM cell voltage of 0.4V, the cell temperatures of the LIB pack at discharge rates of 4C, 6C and 8C are tabulated as observed in Table 4. The temperature gradient at 4C across the HESS in the form of a 3D plot, Fig. 9, shows that the point of maximum temperature is located at the cathode terminal of the PEM fuel cell.

\begin{figure}[htbp]
\centering
{\includegraphics[scale=0.36]{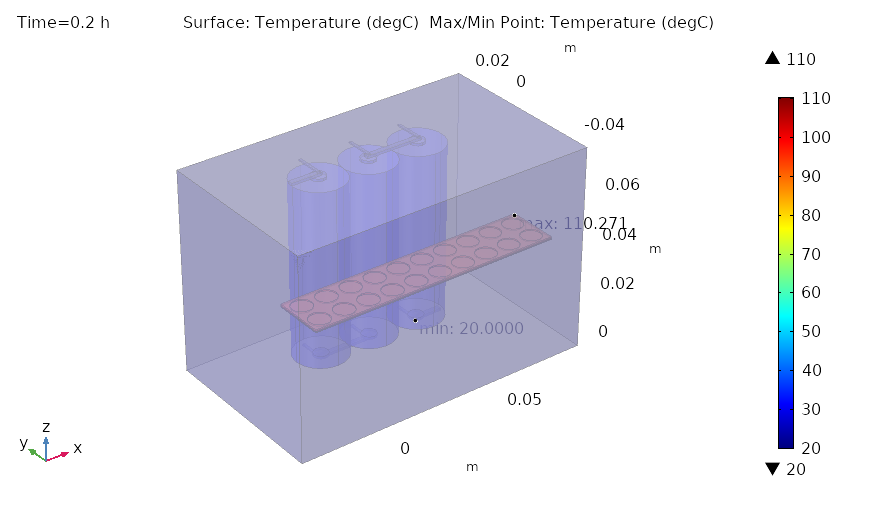}}\\
\caption{Surface Temperature at V\_cell = 0.4V}
\label{fig}
\end{figure}

\begin{table}[htbp]
\caption{CELL TEMPERATURES FOR DIFFERENT C RATES}
\begin{center}
\begin{tabular}{|c|c|c|c|c|}
\hline
\textbf{Time}&\textbf{C}&\multicolumn{3}{|c|}{\textbf{Cell Temperature($^{\circ}$C) }} \\
\cline{3-5} 
\textbf{(hr)} &\textbf{Rate} & \textbf{\textit{Cell 1}}& \textbf{\textit{Cell 2}}& \textbf{\textit{Cell 3}}\\
\hline
\text{0.1} & \text{4C} & \text{26.05} & \text{26.52} & \text{26.05}\\
\hline
\text{0.1} & \text{6C} & \text{31.247} & \text{32.058} & \text{31.247}\\
\hline
\text{0.1} & \text{8C} & \text{35.525} & \text{36.504} & \text{35.525}\\
\hline
\end{tabular}
\label{tab4}
\end{center}
\end{table}

The cell temperature of the LIB pack, when in close proximity to the PEM cell, increases by $\approx$ 0.6$^{\circ}$C – 1.3$^{\circ}$C, with respect to the cell temperature in the absence of a fuel cell, for discharge rates ranging from 4C – 8C, as a result of convective heat transfer from the PEM fuel cell to the LIB pack.

\section{conclusion}

The Hybrid Energy Storage System (HESS) has emerged as a solution to achieve desired performance as it involves the coupling of two or more Energy Storage Systems whose features are combined. The PEM fuel cell and Li-Ion Battery (LIB) hybrid has been immensely used in several applications ranging from hybrid vehicles to unmanned ground vehicles. This paper studies the effect of geometrically placing a LIB pack and a PEM fuel cell in close proximity, as in the cases of a small robotic vehicles or industrial robots which have compact space for on-board power sources. An analysis of the cell temperatures of the LIB pack for discharge rates of 4C, 6C and 8C at different PEM cell voltages is carried out. From the results obtained, it is clearly observed that as the discharge rate of the LIB increases from 4C to 8C and the voltage of the PEM fuel cell decreases from 1V to 0.4V, the cell temperatures of the LIB pack increases. This rise in temperature is a due of convective heat transfer between the storage elements as a result of the compact arrangement of the ESSs. It is necessary to design an efficient cooling system since the individual energy storage systems will exhibit their own thermal behaviour  because of their electrochemical properties , and additional heat from the surrounding environment may cause overheating of the ESSs which in turn may cause damage to the HESS.

\vspace{12pt}


\begin{thebibliography}{00}
\bibitem{b1}	E. L. González, J. S. Cuesta, F. J. V. Fernandez, F. I. Llerena,  Miguel A. R. Carlini, C. Bordons et al., "Experimental evaluation of a passive fuel cell/battery hybrid power system for an unmanned ground vehicle", \textit{International Journal of Hydrogen Energy}, vol. 44, no. 25, 2019, pp. 12772-12782.
\bibitem{b2}	B. Tanç, H. T. Arat, Ç. Conker, E. Baltacioğlu and K. Aydin, "Energy distribution analyses of an additional traction battery on hydrogen fuel cell hybrid electric vehicle", \textit{International Journal of Hydrogen Energy}, 2019.
\bibitem{b3}Y. Wang, Z. Sun, and Z. Chen, "Energy management strategy for battery/supercapacitor/fuel cell hybrid source vehicles based on finite state machine", \textit{Applied Energy}, vol. 254, 2019, p. 113707.
\bibitem{b4}S. Keller, K. Christmann, M. S. Gonzalez and A. Heuer, "A Modular Fuel Cell Battery Hybrid Propulsion System for Powering Small Utility Vehicles," \textit{2017 IEEE Vehicle Power and Propulsion Conference (VPPC)}, Belfort, 2017, pp. 1-4.
\bibitem{b5}Z. Mokrani, D. Rekioua, N. Mebarki, T. Rekioua and S. Bacha, "Energy management of battery-PEM Fuel cells Hybrid energy storage system for electric vehicle," \textit{2016 International Renewable and Sustainable Energy Conference (IRSEC)}, Marrakech, 2016, pp. 985-990.
\bibitem{b6}C. Chao, "Simulation of a fuel cell-battery-ultra capacitor-hybrid-powered electric golf cart," \textit{2019 6th International Conference on Systems and Informatics (ICSAI)}, Shanghai, China, 2019, pp. 1610-1615.
\bibitem{b7}U. Sarma and S. Ganguly, "Determination of Rating Requirement for Fuel-Cell-Battery Hybrid Energy System to Substitute the Diesel Locomotives of Indian Railway," \textit{2017 14th IEEE India Council International Conference (INDICON)}, Roorkee, 2017, pp. 1-6.
\bibitem{b8}C. H. Choi, S. Yu, I. S. Han, B. K. Kho, D. G. Kang, H. Y. Lee et al., "Development and demonstration of PEM fuel-cell-battery hybrid system for propulsion of tourist boat",\textit{ International Journal of Hydrogen Energy}, vol. 41, no. 5, 2016, pp. 3591-3599.
\bibitem{b9}W. R. W. Daud, R. E Rosli, E. H Majlan, S. A. A Hamid, R. Mohamed, and T. Husaini, ”PEM fuel cell system control: A review”, \textit{ Renewable Energy}, vol. 113, 2017, pp. 620-638.
\bibitem{b10}Z. Mokrani, D. Rekioua, N. Mebarki, T. Rekioua and S. Bacha, "Energy management of battery-PEM Fuel cells Hybrid energy storage system for electric vehicle," \textit{2016 International Renewable and Sustainable Energy Conference (IRSEC)}, Marrakech, 2016, pp. 985-990.
\bibitem{b11}W. Chesworth, \textit{Encyclopedia of Soil Science}, Springer Science \& Business Media, 2007.
\bibitem{b12}E. J. Dickinson and G. Hinds, "The Butler-Volmer equation for polymer electrolyte membrane fuel cell (PEMFC) electrode kinetics: A critical discussion", \textit{Journal of The Electrochemical Society}, vol. 166, no. 4, 2019, p. F221.
\bibitem{b13}Shin, Y. Kim, J. Seo, N. Chang, Y. Wang and M. Pedram, "Battery-supercapacitor hybrid system for high-rate pulsed load applications," \textit{2011 Design, Automation \& Test in Europe}, Grenoble, 2011, pp. 1-4.


\end{thebibliography}
\end{document}